\documentclass[conference]{IEEEtran}

\usepackage{graphicx}

\ifCLASSINFOpdf
\else
\fi
\begin{document}
%
\title{Scale Invariant Privacy Preserving Video via Wavelet Decomposition} 
\author{\IEEEauthorblockN{Chengkai Yu\IEEEauthorrefmark{1},
Charles Fleming\IEEEauthorrefmark{2} and 
Hai-Ning Liang\IEEEauthorrefmark{3} }
\IEEEauthorblockA{Department of Computer Science and Software Engineering,
Xi'an Jiaotong Liverpool University\\
Suzhou, China\\
Email: \IEEEauthorrefmark{1}chengkai.yu14@student.xjtlu.edu.cn,
\IEEEauthorrefmark{2}charles.fleming@xjtlu.edu.cn,
\IEEEauthorrefmark{3}haining.liang@xjtlu.edu.cn
}}

\maketitle
%
%

\markboth{Journal of \LaTeX\ Class Files,~Vol.~14, No.~8, August~2015}%
{Shell \MakeLowercase{\textit{et al.}}: Bare Demo of IEEEtran.cls for IEEE Journals}
%



\maketitle

\begin{abstract}
Video surveillance has become ubiquitous in the modern world.  Mobile devices, surveillance cameras, and IoT devices, all can record video that can violate our privacy.  One proposed solution for this is privacy-preserving video, which removes identifying information from the video as it is produced.  Several algorithms for this have been proposed, but all of them suffer from scale issues: in order to sufficiently anonymize near-camera objects, distant objects become unidentifiable.  In this paper, we propose a scale-invariant method, based on wavelet decomposition. This work was originally published as \cite{yu2018scale}.  Please cite the original paper.
\end{abstract}

\begin{IEEEkeywords}
Privacy, anonymization, video
\end{IEEEkeywords}

%
\IEEEpeerreviewmaketitle

\section{INTRODUCTION }

Cameras and camera-embedded devices have become pervasive in our daily
life. Not only wearable devices such as GoPro cameras and Google Glasses
but also surveillance systems, IoT devices, and even drones 
threaten our privacy. First-person videos are especially becoming
very popular among YouTubers and video bloggers. Wearable cameras
are also widely equipped by the police for security and evidence-gathering
purposes. Many such videos are eventually uploaded to the
Internet and processing techniques are often used to remove sensitive
information such as people\textquoteright s faces. However, studies
have shown that video blurring techniques cannot balance privacy
with awareness of risky situations by the person being recorded \cite{Neustaedter2005}.
As the privacy issue of such videos recorded by wearable cameras is
attracting more attention from the public \cite{Doherty2013}, people
are paying more attention to the potential threats to their privacy, and methods to preserve privacy in video need to be developed. W

The usual video processing technique for privacy protection is to anonymize
the video by applying blurring effects on sensitive regions so that the information is
not observable to the viewers. However, common methods for anonymization
processing often require human selection of sensitive regions.
Faces are the only thing that would be blurred out in most 
videos. The effect might not be actually anonymous due to other information
revealed in the videos, the person\textquoteright s body shape, or from
the background and nearby objects. Other methods blur the whole video from, but this has the side effect of making distance too indistinct to separate from the background.

\section{RESEARCH AIM AND OBJECTIVES}

As minor details could as well threaten people\textquoteright s privacy,
actual anonymity using an existing algorithm would in fact, produce a video that is so anonymous
that it destroys most of the visual information, including the objects and background.
However, this would generally make the videos themselves unusable for whatever purpose they were recorded for. Therefore, detecting and performing the blurring effect on different information in the video based on the scale could potentially
blur out the sensitive information but maintain usability. 

\section{LITERATURE REVIEW}

Privacy issues in wearable cameras have been widely discussed with
the growth of the popularity of recording videos by portable devices. Wearable
cameras are more portable than conventional video recording devices
and have emerged as a popular way to capture a wide variety of experiences
that threaten people\textquoteright s privacy more aggressively and
pervasively. Nguyen et al. conducted an extensive study on how individuals
perceive and react to being recorded in first-person videos \cite{Nguyen2009}.
Findings suggest that most people would like to be asked for permission
in case the recordings are shared with others, and most people would
mind if they are being recorded without being notified. 

A study on addressing privacy concerns from videos taken in first-person
point-of-view evaluated the effectiveness of four techniques (face
detection, image cropping, location filtering, and motion filtering)
at reducing privacy infringing content \cite{Thomaz2013}. Minor
information that could be linked back to an individual  as
a non-obvious threat to privacy was pointed out in the study. All
four methods were not particularly effective and could still pose privacy concerns. Glasses-style wearable devices were investigated
with respect to recording and privacy and how these devices differ from other
classes of cameras \cite{Denning2014}. The qualitative study found
that for such subtle devices, reactions to recording can be
affected by the perception of the recorder and whether or not they
could be identified in the recording. It was also pointed out that
people frequently change their perceptions with repeated exposure
or change their views as they become active users of such devices. 

Ethical considerations of wearable cameras were investigated by \cite{Kelly2013}.
Guidelines for all involved and best practices for third parties
were proposed to address the ethical considerations of wearable cameras.
Considerations in pervasive recording technologies were assessed by
\cite{Massimi2010} for an insight into how design, technology, and
policy can work together for the appropriate usage of such technologies.
An interesting finding suggested that a lack of control over recordings
could actually make recording more tolerable. In an inescapable
situation, everything has been decided, and people could rationalize it better than
in controllable situations. Wearable cameras are demonstrated
as an important emerging method to provide personalized feedback and
support in public health interventions \cite{Doherty2013}. Ethical
approval and privacy concerns are the most significant barriers
and require more research in this domain. An important goal is to
create interventions that explain, control, and notify people about
the technologies. 

The issue of privacy protection has been widely discussed in computer
vision, especially when it comes to human recognition from sources
such as robot \cite{Butler2015}, wearable cameras \cite{Templeman2014,Poleg2014},
first-person video \cite{Narayan2014}, or surveillances videos \cite{Niu2004}.
Human activity recognition has received a great amount of attention
and recognition algorithms for various environments and activities
were introduced \cite{Ryoo2016,Dai2015,Aggarwal2011} and \cite{zhang2021multi} which highlight
the importance of privacy-preserving videos. However, studies that
cover privacy issues were focused on approaches for protecting peoples\textquoteright{}
privacy but not on analyzing the actual privacy protection of the
current anonymous algorithms. Privacy is a key factor in adopting new technologies \cite{liang2020personal} \cite{olade2019smart}.

While some attempts to use adversarial attacks such as \cite{cilloni2022ulixes} have been proposed, they do not fully address the issue. Systems to prevent leakage, such as \cite{fleming2012data}, also do not address the problem when the data is willingly uploaded.

\section{ANONYMIZATIONS ALGORITHM}

As baseline algorithms, we compare with three commonly used blurring algorithms, including Gaussian blur, down-sampling, and superpixel, applying these to our test videos. The three algorithms tend
to produce different blurry effects and thus were chosen for determining
and assessing their degree of anonymization in surveillance and
wearable settings. We compare our algorithm versus this standard algorithm on both up close and distant objects in videos and show that our algorithm outperforms all three.

\section{Wavelet Transformation }

A wavelet transformation decomposes the signal into several bands
to capture and separate different characteristics of the original
signal. The signal information of an object contains detectable differences, which could be captured by one or more bands during wavelet decomposition.
The edges and surfaces are separated by different sets of bands
due to the different characteristics of their signal. Different levels
tend to capture the changes in color or textures in original images. 

The wavelet transformation technique used in the study is the discrete
wavelet transformation (DWT). The discrete wavelet transformation
of an image signal of a level is calculated by passing it into
a filter bank, where a series of filters perform levels of decomposition.  The
signal is transformed by a high-pass filter which gives the output
of the detail coefficients, while the low-pass filter returns an approximation
coefficient. As the image signal has been decomposed at the current
level, the output of the low-pass filter will then be subsampled by
2 in the next level. 

The Wavelet transformation anonymization algorithm (WTAA) demonstrates
the potential of balancing anonymity and usage by performing scalable
blurring effects based on the scale of the objects. We anonymize the video by decomposing the video using the wavelet transform, selectively destroying certain wavelet coefficients, then reconstructing the image via the inverse transform. 

We compare the performance of our algorithm by considering a series of videos with both near and far objects.  The wavelet transformation anonymization algorithm shows a
great improvement when objects are far away. Gaussian
blur has good anonymization but preserves insufficient
information for usage. Downsampling has the lowest level of anonymization
and does not keep the outline of the object from a distance. Superpixel
has a reasonable anonymization level and preserves part of the shape;
however, would not be sufficient to be considered useful. Wavelet
transformation can keep both the shape and the color at a comparable
level to the original video and performs extremely well in preserving
distance objects due to the inherent scale in the wavelet decomposition. 

\begin{figure*}
\begin{centering}
\includegraphics[width=5in]{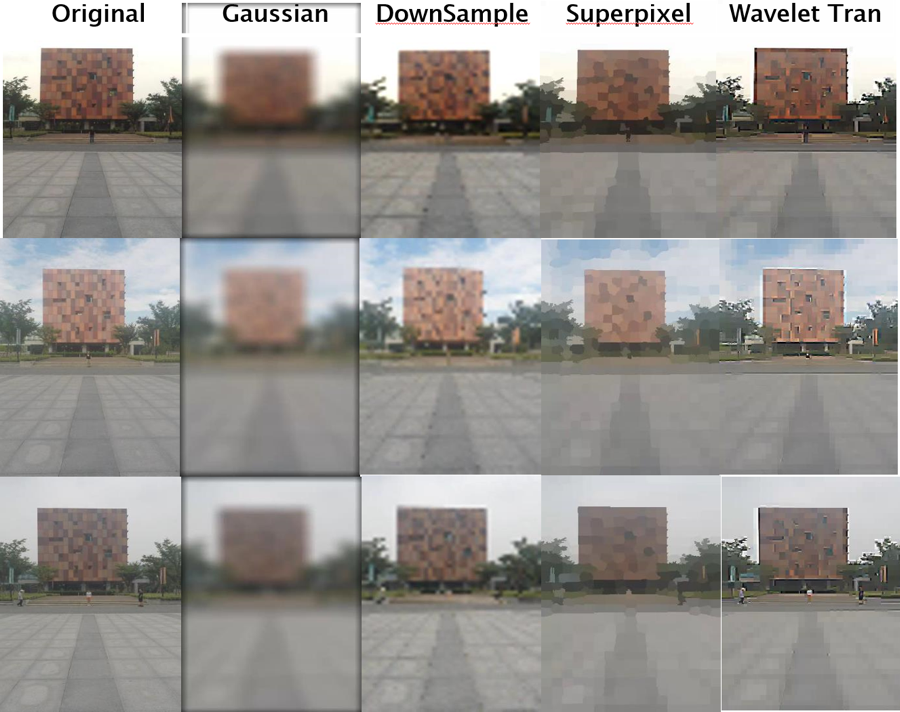}
\par\end{centering}
\centering{}\caption{Comparison of anonymization algorithms for distant figures. }
\end{figure*}

A the zoomed-in shot, the person is about 15 meters away from the camera. Gaussian blur does not preserve any shape or color at this
distance in order to maintain the level of anonymity. Downsampling
preserves a degree of color when the color contrast to the background
is large, but the shape is left out. Superpixel merges the person
with adjacent superpixels, and the color would be destroyed as well as
part of the shape. WTAA 
preserves both the shape and the color enough for the figure to be recognizable as a person. This effect is more pronounced when viewed as a video.

\begin{figure*}
\begin{centering}
\includegraphics[width=5in]{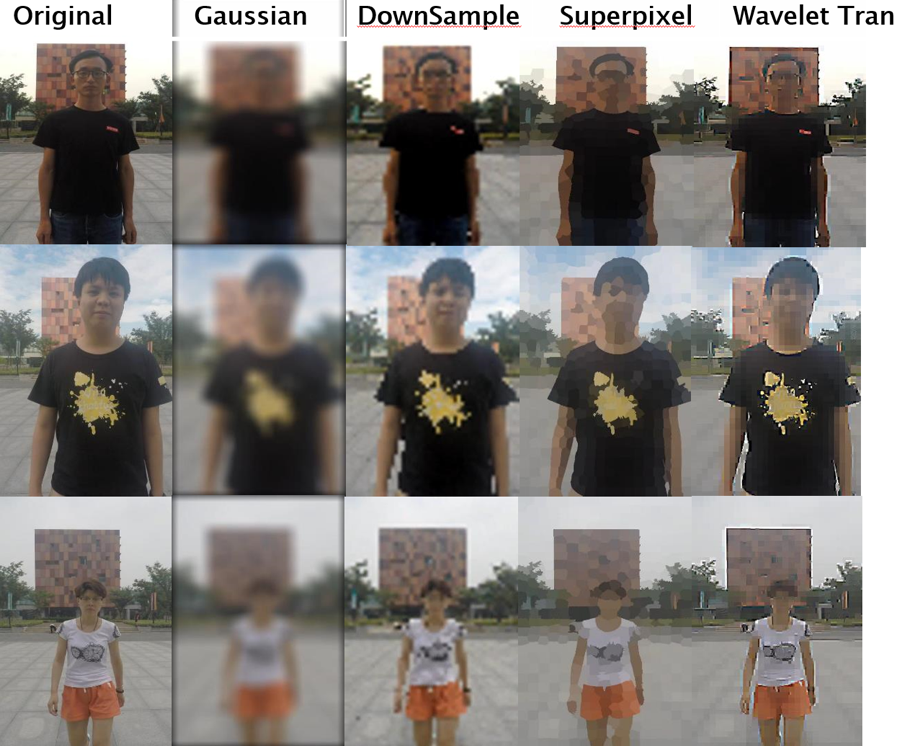}
\par\end{centering}
\centering{}\caption{Comparison in the level of anonymity in the last frame.}
\end{figure*}

\bibliographystyle{IEEEtran}
\bibliography{localBib}

\end{document}